D.S. Schmerling (Higher School of Economics, National Research University).

**New models of income distribution – "graduation"**

**as the explanation of Gini coefficient (index).**

**Annotation**
The paper covers the new model of wage distribution in typical group of people. The model provides the opportunity to reparameterize applicable income distribution model (Pareto, logarithmically normal, logarithmically logistic, Dagum's etc.). The model ensures "the graduation" of Gini index values by polynomial degree of wage distribution as well as different types of income distribution. The given approach clarifies the nature of income inequality.

**Summary**

The work covered the topical problem as the study of income and wealth inequality introduces the notion of "graduation" i.e. rate of labour grade as it is accepted at time wage. The graduation gives the opportunity to introduce the index m as degree of polynomial on the basis of which the scale of wage rates is distributed and as the result the fact can allow to calculate and estimate Gini index. The last one gives the opportunity to graduate Gini index from 0 to 1 assigning integral number or fraction *m, 0, <m<∞ to* every value of Gini index. Then these known income distributions are subject to graduation (Pareto, logarithmically logistic, logarithmically normal). All above mentioned clarifies wide-spread Gini coeffient to some extend and "the nature of populations' wealth".

Income distribution problem has been known at least since M. Lorenz's paper (Lorenz M.O., 1905)[1] about concentration of wealth, e.g. refer to Kendall, Stuart, Ord (1977-1983), §§ 2.25[2].

Wide range of literature on the problem of income, wealth and personal property differentiation of population, enterprises and territory exist.

Income distribution problem is widely used by Gini coefficient (index) (Corrado Gini, 1912).

Let

---

[1] Biography of M. Lorenz, see Marshall, Olkin (1980), refer to Lorenz_curve.Wikipedia.html.
[2] Besides, inequality was mentioned in V.I.Lenin's paper «Development of capitalism in Russia» (1899).

$$\Delta_1 = \frac{1}{N(N-1)} \sum_{j=-\infty}^{\infty}\sum_{k=-\infty}^{\infty} |x_j - x_k| f(x_j) f(x_k) \quad (1)$$

(for discrete distribution),

where $x_1, x_2, \ldots$ - value of income, $f(x_1), f(x_2)$ – probability (frequency) of people with income $x_1, x_2, \ldots$ respectively.

The normalized value $\Delta_1$ is equal to $\Delta_1^* = \Delta_1/\Delta_1^{max}$ € [0,1], means the greater social inequality.

Besides the area over Lorenz curve and under diagonal (picture 2.2 Kendall, Stuart, Ord, 1977-1983) is equal to $\Delta_1/4 \mu_1$, where $\mu_1 = \int_{-\infty}^{\infty}(x-a)f(x)dx$,

$f(x)$ – distribution density, assigned a=0, (§§ 2,3 Kendall, Stuart, Ord, 1977-1983).

Dissipation curve is equal to

$$\phi(x) = \frac{1}{\mu_1'} \int_{-\infty}^{x} xf(x)dx \quad (2)$$

i.e. «incomplete first distribution's moment» (the same reference).

The mean difference computation (the same reference).

Gini coefficient has been calculated and published for most countries for decades, the methods of evaluation and grouped data as well see Gastwirth (1972), Modorres, Gastwirth (2006)[3].

Some value of normalized Gini index see in the table 1[4] (due to CIA data), for example in Norway 0,25 (2008), France 0,32 (2008), Russia 0,423 (2008), Nigeria 0,437 (2003), the USA 0,45 (2007), Mexico 0,482 (2008), Haiti 0,538 (2001), Sierra Leone 0,629 (1989), South Africa 0,65 (2005), Namibia 0,707 (2003)[5], and later is up to 0,75.

The reader-economist is probably used to the data but the question is how we understand Gini coefficient?

The wide range of literature exists about the negative effect of high Gini coefficient (> 0,3), refer to Handbook of income distribution (Atkinson, Bourguignon, 2000). The value 0,3 or 1/3 (0,33) is arbitrary enough and can be equal to average European Gini coefficient.

It would be useful to assign to the normalized coefficient G', 0 ≤ G' ≤ 1, some economical interpretation.

---

[3] See Morgan J. The Anatomy of Income Distribution // Rev.Econ.Statist. (MITpress), 1962, v.44, N 3, p. 270-283.
[4] Article in Wikipedia…List of Countries by income equality
[5] In Bolivia Gini index is 0,592, but decile rate is too high – 168, 1 (!)

Let's consider the following model. Let $x_i$ – income of people referring to i level of hierarchy, concerning enterprise, population, territory, etc., i = 1, ..., n. Model P ("graduation") is like that.

Income at i level (i = 1 – population with the lowest income, and i = n – c with the highest income) is equal to $x_i$ = Const $i^m$, m = 1, 2, 3, ..., M > 0.

<u>Theorem</u>. Gini coefficient (index) $G_m'(n)$ for the model P is equal to asymptotically at n → ∞

$$G_m'(n) \to \frac{m}{m+2} \quad (1a)$$

<u>Theorem proving</u>

$$G'(n) = \frac{"\sigma"(n)}{\max_x "\sigma"(n)}, \quad (2a)$$

where maximum for every possible $\{x_{(1)}, x_{(2)}, ..., x_{(n)}\}$, so

$$\sum_{1 \le i \le n} x_{(i)} = C(n), \quad (3)$$

$$"\sigma"(n) = \frac{2\sqrt{\pi}}{n(n-1)} \sum_{1 \le i \le n} (i - \frac{n+1}{2}) x_{(i)}, \quad (4)$$

$X_{(i)}$ – i-order statistics[6]

$$G(n) = \frac{1}{n(n-1)} \sum_{1 \le i, j \le n} |x_i - x_j|. \quad (6)$$

Using another form "σ"(n) = "σ"

---

[6] See David (1981), (7.4.1.) and §7.4., ex. 7.4.1., §9.6, (9.6.1.), where asymptotical "σ" is discussed.

So $E"\sigma" = 2\sqrt{\pi \int_{-\infty}^{+\infty} x \left[ P_{(x)} - \frac{1}{2} \right] dP_{(x)}}$, "σ" – unbiased estimator for normal samples, $P_{(x)}$ – cumulative distribution function.

$$"\sigma" = \frac{\tilde{\kappa} 2 * \sqrt{\pi}}{n(n-1)} \left\{ \sum_1^n i\, x_{(i)} - \frac{n+1}{2} \sum_1^n x_{(i)} \right\} \quad (7)$$

and calculating

$$\max_x "\sigma" = \frac{\sqrt{\pi} * \tilde{\kappa}}{n} S_m(n) \quad (8)$$

where $S_m(n) = \sum_{k=0}^{n-1} k^m$ [7]

the sum of integers raising to power m=1, 2, 3,…, k = 0, 1, 2,…., n-1, allows us to have a formula for (3).

So let

$$"\sigma" = \frac{2\tilde{\kappa}\sqrt{\pi}}{n(n-1)} \left\{ \tilde{S}_{m+1}(n) - \frac{n+1}{2} \tilde{S}_m(n) \right\} / \frac{\tilde{\kappa}\sqrt{\pi}}{n} \tilde{S}_m(n) \quad (9)$$

Therefore (here m means degree of polynomial)

$$G'_m(n) = \frac{2}{n-1} \left\{ \frac{\tilde{S}_{m+1}(n)}{\tilde{S}_m(n)} - \frac{n+1}{2} \right\} \quad (10)$$

According to Graham, Knuth, Patashnik (1994), (61.78),

$$\tilde{S}_m(n) = \frac{1}{m+1} \sum_{0 \le k \le n} \binom{m+1}{k} B_k n^{m+1-k} \quad (11),$$

where $B_k$, k = 0, 1, 2,…., Bernoulli's number (Jacob), that is

Table 1

| k | 1 | 2 | 3 | 4 | 5 | 6 | 7 | 8 | 9 | 10 | 11 | 12 | … |
|---|---|---|---|---|---|---|---|---|---|---|---|---|---|
| $B_k$ | 1 | $-\frac{1}{2}$ | 0 | $-\frac{1}{30}$ | 0 | $\frac{1}{42}$ | 0 | $-\frac{1}{30}$ | 0 | $\frac{5}{66}$ | 0 | $-\frac{691}{2730}$ | … |

---

[7] See Graham, Knuth, Patashnik (1994).

Formulae Sm(n), m = 0, 1, 2,…, 10 see the above mentioned book p.314.

Formula (10) implies at n→∞

$$G'_m(n) \cong \frac{2}{n-1}\left\{\frac{m+1}{m+2}n - \frac{n+1}{2}\right\} \approx \frac{m}{m+2}, QED.$$

for $G'_m(n)$ [8]

Table 2

| m | 1 | 2 | 3 | 4 | 5 | 6 | 7 | 8 | 9 | 10 | … |
|---|---|---|---|---|---|---|---|---|---|----|---|
| $G'_m(n)$ | $\frac{1}{3}$ | $\frac{1}{2}$ | $\frac{3}{5}$ | $\frac{2}{3}$ | $\frac{5}{7}$ | $\frac{3}{4}$ | $\frac{7}{9}$ | $\frac{4}{5}$ | $\frac{9}{11}$ | $\frac{5}{6}$ | … |

Let's revert to the beginning of the summary. If P is equal to Const = 1, so $G'$=0,25 (Norway) means m < 1 (fraction), $G'$=0,327 (France) means m≈1, $G'$=0.423 (Russia) means 1<m<2, $G'$=0,482 (Mexico) provides m≈2, $G'$=0.538 (Haiti) – 2<m<3, Sierra Leone with 0,629 leads to 3<m<4, Republic of South Africa with 0,65 provides m more, but Namibia, according to different sources, 0,707< $G'$ <0.75, can have pretentions to 5<m<6.

Let us clarify our results.

1) Leopold Kronecker (1823-1891) said that God invented integer numbers, and people do all the rest.

2) Companies, communities, regions and countries can be relatively divided into linear (m=1), quadratic (m=2), cubic (m=3), «tetradic» (m=4), «pental-power» (m=5), «hexal» (m=6) etc.

For example, for Moscow, where $G'$ raised 0,62 in different years it can be assumed cubic type of income distribution. But we should keep in mind that the true value of $G'$ in Moscow can be equal to 0.60-0.70 according to most specialists.

3) As for Russian companies the data of some large companies which are regularly discussed by "Vedomosti" show the following results[9]. The bonuses of administration personnel (exclusive of salary) compiles 1-3 million USD per year compared to salary of senior staff at the level i=1 compiles 1.5-3.0 thousand USD and bonus which is equal to 10 – 30 thousand USD altogether form the value 2<m<3. m can probably reach 4, as in 2010 the General Manager's salary of the leading companies in Russia can reach 25 million USD per year. See htm://www.rfcor.ru/print/news_rfc_987htm.

Ratio m "graduate" (the notion is taken from physics and engineering) "Gini scale".

---

[8] At m=1 - Gini index is accurate (definite), at every n = 1, 2, 3….
[9] See magister dissertation: Milek O. The study of income distribution with the help of heavy tail distribution. – M., HSE, 2010 (in Russian)

So the question can arise if the ways of income distribution reasonable enough at different G'?  See  Atkinson, Bourguignion (2000), Chacravarty (1990), Foster, Sen (1997).

Kiruta A. Ya. analyzed the discussed here problem and sent the 8-paged comment to the author on May, 3rd, 2011, at 2.34 pm.

**It should be mentioned that it is also possible to consider real values (due to interpolation)**

$$m=2G^{\wedge}/(1-G^{\wedge}).$$

**Here  G^ - normalized value of Gini index.**

**For Moscow in 2009 G^=0.521, that gives m=2.742.**

For Norway (2008) G^=0.25 and m=0.667.

Here clarification should be done.

The above discussed model can be correlated to traditional static distribution. For Pareto distribution with the same G^, as in our model P,  m < 1, ensures finite variance (dispersion), but for log-logistic distribution m < 2 is required. As for log-normal distribution variance (dispersion) doesn't approach infinity, but it is increasing slowly on condition that m is slowly increasing as well. It should be mentioned that Pareto and log-logistic distributions describe right (top) tail but log-normal ones - little income and describe poorly right tail.

The summary is that at high degree on inequality in the model P little (rich) part of the society strives to increase the income so that the high distribution tails are getting heavy and variance (dispersion) tends to infinity. At the same time the part of the society with average income reacts slowly at the increase of m.

 The author of the article was inspired by the report of Vladimir I. Arnold "Hard" and "soft" mathematical models" presented on September, 25the, 1997, in Administration of the President of the Russian Federation at seminar "Analytics in State Institutions" under the guidance of Rajkov A.N., Satarov G.A., Schmerling D.S.

The author thanks Arnold V.I,. Kiruta A. Ya., Nikitin Ya. Yu., Oberemko A.I., Orlov A.I., Tolstova Yu. N., Tyurin Yu. N., Uljanov V.V. for assistance and discussion.